\documentclass[%
 aip,
cp,  
 amsmath,amssymb,
 reprint,%
]{revtex4-2}

\usepackage{graphicx}
\usepackage{dcolumn}
\usepackage{bm}

\usepackage[utf8]{inputenc}
\usepackage[T1]{fontenc}
\usepackage{mathptmx}

\usepackage{color}

\usepackage[mathscr,scaled=1.15]{urwchancal}
\DeclareFontFamily{OT1}{pzc}{}
\DeclareFontShape{OT1}{pzc}{m}{it}%
{<-> s * [1.15] pzcmi7t}{}
\DeclareMathAlphabet{\mathpzc}{OT1}{pzc}{m}{it}

\definecolor{olive}{rgb}{0.3, 0.4, .1}
\definecolor{fore}{RGB}{249,242,215}
\definecolor{back}{RGB}{51,51,51}
\definecolor{title}{RGB}{255,0,90}
\definecolor{dgreen}{rgb}{0.,0.6,0.}
\definecolor{gold}{rgb}{1.,0.84,0.}
\definecolor{JungleGreen}{cmyk}{0.99,0,0.52,0}
\definecolor{BlueGreen}{cmyk}{0.85,0,0.33,0}
\definecolor{RawSienna}{cmyk}{0,0.72,1.00,0.85}
\definecolor{Magenta}{cmyk}{0,1,0,0}
\definecolor{DarkerRed}{rgb}{1.00,0.10,0.10}
\definecolor{DarkerFuchsia}{RGB}{118,0,118}

\usepackage{titlesec}
\titlespacing*{\section}{0pt}{0.45\baselineskip}{0.45\baselineskip}
\titlespacing*{\subsection}{0pt}{0.45\baselineskip}{0.45\baselineskip}

\begin{document}

\title{\hspace*{-0.60cm} Nucleon-to-Resonance Form Factors at Large Photon Virtualities}

\author{J. Segovia} 
\email[Corresponding author: ]{jsegovia@upo.es}
\affiliation{Departamento de Sistemas F\'isicos, Qu\'imicos y Naturales, Universidad Pablo de Olavide, E-41013 Sevilla, Spain}

\author{C. Chen}
\email{Chen.Chen@theo.physik.uni-giessen.de}
\affiliation{Institut f\"ur Theoretische Physik, Justus-Liebig-Universit\"at Gie\ss en, Heinrich-Buff-Ring 16, 35392 Giessen, Germany}

\author{Z.-F. Cui}
\email{phycui@nju.edu.cn}
\affiliation{Department of Physics, Nanjing University, Nanjing, Jiangsu 210093, China}

\author{Y. Lu}
\email{luya@nju.edu.cn}
\affiliation{Department of Physics, Nanjing University, Nanjing, Jiangsu 210093, China}

\author{C.D. Roberts}
\email{cdroberts@anl.gov}
\affiliation{Physics Division, Argonne National Laboratory, Lemont, Illinois 60439, USA}


\date{\today} 

\begin{abstract}
We present a unified description of elastic and transition form factors involving the nucleon and its resonances; in particular, the $N(1440)$, $\Delta(1232)$ and $\Delta(1600)$. We compare predictions made using a framework built upon a Faddeev equation kernel and interaction vertices that possess QCD-kindred momentum dependence with results obtained using a confining, symmetry-preserving treatment of a vector$\,\otimes\,$vector contact-interaction in a widely-used leading-order (rainbow-ladder) truncation of QCD's Dyson-Schwinger equations. This comparison explains that the contact-interaction framework produces hard form factors, curtails some quark orbital angular momentum correlations within a baryon, and suppresses two-loop diagrams in the elastic and transition electromagnetic currents. Such defects are rectified in our QCD-kindred framework and, by contrasting the results obtained for the same observables in both theoretical schemes, shows those objects which are most sensitive to the momentum dependence of elementary quantities in QCD.
\end{abstract}

\maketitle


\section{Introduction}

A unified description of electromagnetic elastic and transition form factors involving the nucleon and its resonances has acquired very much interest. On the theoretical side, it is via the $Q^2$-evolution of form factors that one gains access to the running of QCD's coupling and masses~\cite{Cloet:2013gva, Chang:2013nia}. Moreover, QCD-kindred approaches that compute form factors at large photon virtualities are needed because the so-called meson-cloud screens the dressed-quark core of all baryons at low momenta~\cite{Tiator:2003uu, Kamano:2013iva, Mokeev:2015lda}. On the experimental side, substantial progress has been made in the extraction of transition electrocouplings, $g_{{\rm v}NN^\ast}$, from meson electroproduction data, obtained primarily with the CLAS detector at the Jefferson Laboratory (JLab)~\cite{Aznauryan:2011qj, Mokeev:2018zxt, Isupov:2017lnd, Mokeev:2012vsa, Markov:2019fjy}. The electrocouplings of all low-lying $N^\ast$ have been determined via independent analyses of $\pi^+ n$, $\pi^0p$ and $\pi^+ \pi^- p$ exclusive channels~\cite{Aznauryan:2009mx, Mokeev:2012vsa, Mokeev:2015lda, Tanabashi:2018oca}; and preliminary results for the $g_{{\rm v}NN^\ast}$ of some high-lying $N^\ast$ states, with masses below $1.8\,{\rm GeV}$, have also been obtained from CLAS meson electroproduction data~\cite{Aznauryan:2012ba, Mokeev:2018zxt}. Full up-to-date information on $Q^2$ evolution of $g_{{\rm v}NN^\ast}$ electro-couplings at $Q^2<6.0\,\text{GeV}^2$ for most resonances in the mass range up to $1.8\,\text{GeV}$ from analyses of exclusive meson electro-production with CLAS can be found in Ref.~\cite{Blin:2019fre}.

During the next decade, CLAS\,12 will deliver resonance electroproduction data out to $Q^2 \approx 12\,$GeV$^2$~\cite{Mokeev:2018zxt, Burkert:2018nvj, Burkert:2019opk} and thereby empirical information which can address a wide range of issues that are critical to our understanding of strong interactions, \emph{e.g}.: is there an environment sensitivity of DCSB; and are quark-quark correlations an essential element in the structure of all baryons?  Existing experiment-theory feedback suggests that there is no environment sensitivity for the $N(940)$, $N(1440)$ and $\Delta(1232)$ baryons: DCSB in these systems is expressed in ways that can readily be predicted once its manifestation is understood in the pion, and this includes the generation of diquark correlations with the same character in each of these baryons. Resonances in other channels, however, probably contain additional diquark correlations, with different quantum numbers, and can potentially be influenced in new ways by meson-baryon final state interactions (MB\,FSIs).  Therefore, these channels, and higher excitations, open new windows on nonperturbative QCD and its emergent phenomena whose vistas must be explored and mapped if the most difficult part of the Standard Model is finally to be solved.

This manuscript is arranged as follows. We present in Sec.~II a short survey of our theoretical framework in order to compute the mass and wave function of the nucleon and its resonances. Section~III is dedicated to the calculation, within the same formalism, of the nucleon's elastic and N$\to$Roper form factors. We discuss in Sec.~IV the $\gamma^{\ast}p\to \Delta(1232)^+,\,\Delta(1600)^+$ reactions and their related form factors. We summarize and give some conclusions in Sec.~V.


\section{\label{sec:wave} Baryon wave function}

A baryon is described in quantum field theory by a Faddeev amplitude, obtained from a Poincar\'e-covariant Faddeev equation, which sums all possible quantum field theoretical exchanges and interactions that can take place between the three dressed-quarks that characterise its valence-quark content.

\begin{figure}[!t]
\centerline{%
\includegraphics[clip, height= 0.15\textheight, width=0.70\textwidth]{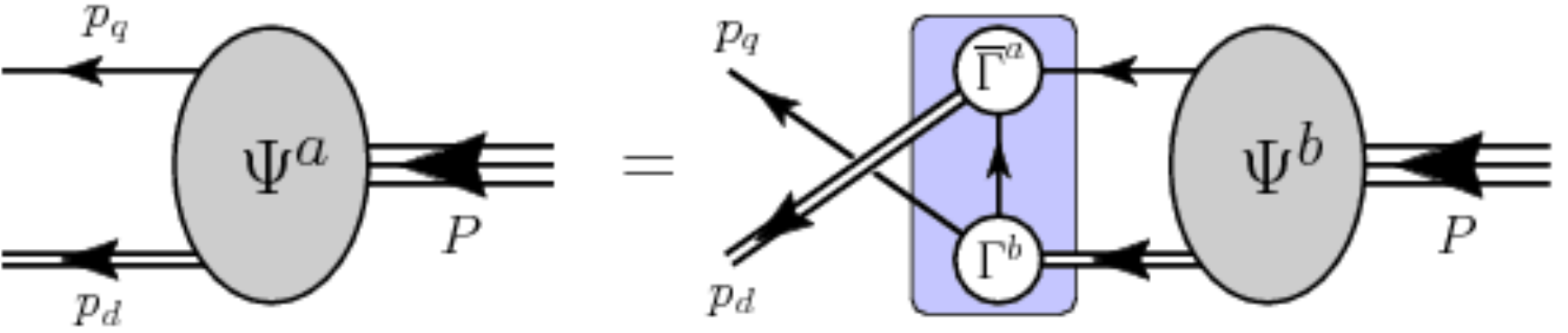}}
\caption{\label{figFaddeev} Faddeev equation: a linear integral equation for the matrix-valued function $\Psi$, being the Faddeev amplitude for a baryon of total momentum $P = p_q + p_d$, which expresses the relative momentum correlation between the dressed-quarks and -nonpointlike-diquarks within the baryon. The shaded rectangle demarcates the kernel of the Faddeev equation: \emph{single line}, dressed-quark propagator; $\Gamma$,  diquark correlation amplitude; and \emph{double line}, diquark propagator.}
\vspace*{-0.40cm}
\end{figure}

A dynamical prediction of Faddeev equation studies that employ realistic quark-quark interactions~\cite{Binosi:2014aea, Binosi:2016nme} is the appearance of non-pointlike quark$+$quark (diquark) correlations within baryons, whose characteristics are determined by DCSB~\cite{Maris:2002yu, Maris:2004bp}. Consequently, the baryon bound-state problem is transformed into solving the linear, homogeneous matrix equation depicted in Fig.~\ref{figFaddeev}~\cite{Cahill:1988dx, Burden:1988dt, Reinhardt:1989rw, Efimov:1990uz}. Its key elements are the dressed-quark and -diquark propagators, and the diquark Bethe-Salpeter amplitudes.

Evidence supporting the presence of diquark correlations in baryons is accumulating; see, for instance, Refs.~\cite{Eichmann:2009qa, Cates:2011pz, Segovia:2013rca, Roberts:2013mja, Segovia:2014aza, Segovia:2015ufa, Segovia:2016zyc, Eichmann:2016hgl, Eichmann:2016yit, Lu:2017cln, Chen:2017pse, Mezrag:2017znp, Chen:2018nsg, Chen:2019fzn}. It should be emphasised that these correlations are fully dynamical and appear in a Faddeev kernel which requires their continual breakup and reformation. Consequently, they are vastly different from the static, pointlike diquarks introduced originally~\cite{Anselmino:1992vg} in an attempt to solve the so-called ``missing resonance'' problem~\cite{Aznauryan:2011qj}. In fact, consistent with numerical simulations of lattice-QCD~\cite{Edwards:2011jj}, the spectrum of states produced by the Faddeev equation in Fig.~\ref{figFaddeev} possesses a richness that cannot be explained by a two-body model.

With the inputs drawn from Refs.~\cite{Segovia:2014aza, Chen:2019fzn} (including light-quark scalar and axial-vector diquark masses $m_{0^+}=0.79\,$GeV, $m_{1^+} = 0.89\,$GeV, respectively) one can readily construct the relevant Faddeev equation kernels of Fig.~\ref{figFaddeev} and use \emph{ARPACK} software to obtain the mass and Faddeev amplitude of the two lightest states in the  $(I,J^P)=(1/2,1/2^+)$ and $(I,J^P)=(3/2,3/2^+)$ channels; which we identify with the baryons $N(940)$, $N(1440)$, $\Delta(1232)$ and $\Delta(1600)$, respectively. The masses are (in GeV):
\begin{equation}
\label{eqMasses}
\begin{array}{cccc}
m_{N(940)} \quad\quad & \quad\quad m_{N(1440)} \quad\quad & \quad\quad m_{\Delta(1232)} \quad\quad & \quad\quad m_{\Delta(1600)} \\
1.19 & $1.73$ & 1.35 & 1.79
\end{array}\,.
\end{equation}
These values correspond to the locations of the two lowest-magnitude poles in the three-quark scattering problems in the given channels. The residues associated with these poles are the Poincar\'e-covariant wave functions, $\chi(\ell^2,\ell\cdot P; P^2)$, where $\ell$ is the quark-diquark relative momentum. For every baryon considered herein, eight scalar functions are required to completely describe the system, each associated with a particular Dirac-matrix structure.

It is important to emphasize that our Faddeev equation should be understood as producing the dressed-quark core of the bound-state and not the completely-dressed object. In other words, a deeper consideration of the kernel in Fig.~\ref{figFaddeev} reveals that resonant contributions, \emph{viz}.\ meson-baryon final-state-interactions (MB\,FSIs), are omitted~\cite{Eichmann:2008ae, Eichmann:2008ef}. Clothing the nucleon's dressed-quark core by including resonant contributions to the kernel produces a physical nucleon whose mass is $\approx 0.2\,$GeV lower than that of the core~\cite{Ishii:1998tw, Hecht:2002ej}. Similarly, MB\,FSIs reduce the $\Delta(1232)$-baryon's core mass by $\approx 0.16\,$GeV~\cite{JuliaDiaz:2007kz, JuliaDiaz:2007fa, Suzuki:2009nj} and the Roper resonance's core-mass by $0.3\,$GeV~\cite{Suzuki:2009nj}. Evidently, such reductions shift the mass of a given baryon's dressed-quark core into alignment with the measured Breit-Wigner mass of the associated physical state.  Moreover, this pattern is seen to prevail broadly, extending to baryons in the multiplets of flavour-$SU(5)$~\cite{Qin:2019hgk, Yin:2019bxe}.

\newpage

\section{\label{sec:nucleon} THE $\gamma^{\ast}p\to N(940),\,N(1440)$ TRANSITIONS}

The ground-state neutron and proton (nucleons) are certainly bound-states seeded by three valence-quarks: $udd$ and $uud$, respectively. However, the nature of the nucleon's first excited state -- $N(1440)\,1/2^+$ -- is less certain. The $N(1440)\,1/2^+$ ``Roper resonance'' was discovered in 1963~\cite{Roper:1964zza, BAREYRE1964137, AUVIL196476, PhysRevLett.13.555, PhysRev.138.B190}, but it was immediately a source of puzzlement because, \emph{e.g}.\ a wide array of constituent-quark potential models produce a spectrum in which the second positive-parity state in the baryon spectrum lies above the first negative-parity state~\cite{Capstick:2000qj, Crede:2013sze}.

Following the acquisition and analysis of a vast amount of high-precision nucleon-to-resonance electro-production data with single- and double-pion final states on a large kinematic domain of energy and momentum-transfer, development of a sophisticated dynamical reaction theory capable of simultaneously describing all partial waves extracted from available, reliable data, and formulation and wide-ranging application of a Poincar\'e covariant approach to the continuum bound state problem in relativistic quantum field theory, it is now widely accepted that the Roper is, at heart, the first radial excitation of the nucleon, consisting of a well-defined dressed-quark core that is augmented by a meson cloud, which both reduces the Roper's core mass by approximately 20\% and contributes materially to the electro-production transition form factors at low-$Q^2$~\cite{Golli:2017nid, Burkert:2019djo}.

We are going to review herein the calculation, consistent with the formalism presented above to the baryon bound-state problem, of the nucleon's elastic form factors and the so-called equivalent Dirac and Pauli form factors of the $\gamma^{\ast}N(940)\to N(1440)$ reaction. This section is mostly based on the work presented in Refs.~\cite{Segovia:2014aza, Segovia:2015hra, Segovia:2015ufa, Segovia:2016zyc, Mezrag:2017znp, Chen:2017pse, Chen:2018nsg, Chen:2019fzn, Segovia:2019nzm}.


\subsection{Transition Current}

The computation of the desired elastic and transition form factors is a straightforward numerical exercise once the Faddeev amplitudes for the participating states are in hand and the electromagnetic current is specified.  When the initial and final states are $I=1/2$, $J=1/2^+$ baryons, the current is completely determined by two form factors, \emph{viz}.
\begin{equation}
\bar u_{f}(P_f)\big[ \gamma_\mu^T F_{1}^{fi}(Q^2)+\frac{1}{m_{{fi}}} \sigma_{\mu\nu} Q_\nu F_{2}^{fi}(Q^2)\big] u_{i}(P_i)\,,
\label{NRcurrents}
\end{equation}
where: $u_{i}$, $\bar u_{f}$ are, respectively, Dirac spinors describing the incoming/outgoing baryons, with four-momenta $P_{i,f}$ and masses $m_{i,f}$ so that $P_{i,f}^2=-m_{i,f}^2$; $Q=P_f-P_i$; $m_{{fi}} = (m_f+m_{i})$; and $\gamma^T \cdot Q= 0$.

The vertex sufficient to express the interaction of a photon with a baryon generated by the Faddeev equation in Fig.~\ref{figFaddeev} is described elsewhere~\cite{Oettel:1999gc, Segovia:2014aza}. It is a sum of six terms, depicted in the Appendix~C of Ref.~\cite{Segovia:2014aza}, with the photon probing separately the quarks and diquarks in various ways, so that diverse features of quark dressing and the quark-quark correlations all play a role in determining the form factors.  


\subsection{Form Factors}

\begin{figure}[!t]
\begin{center}
\begin{tabular}{ccc}
\includegraphics[clip, width=0.45\linewidth]{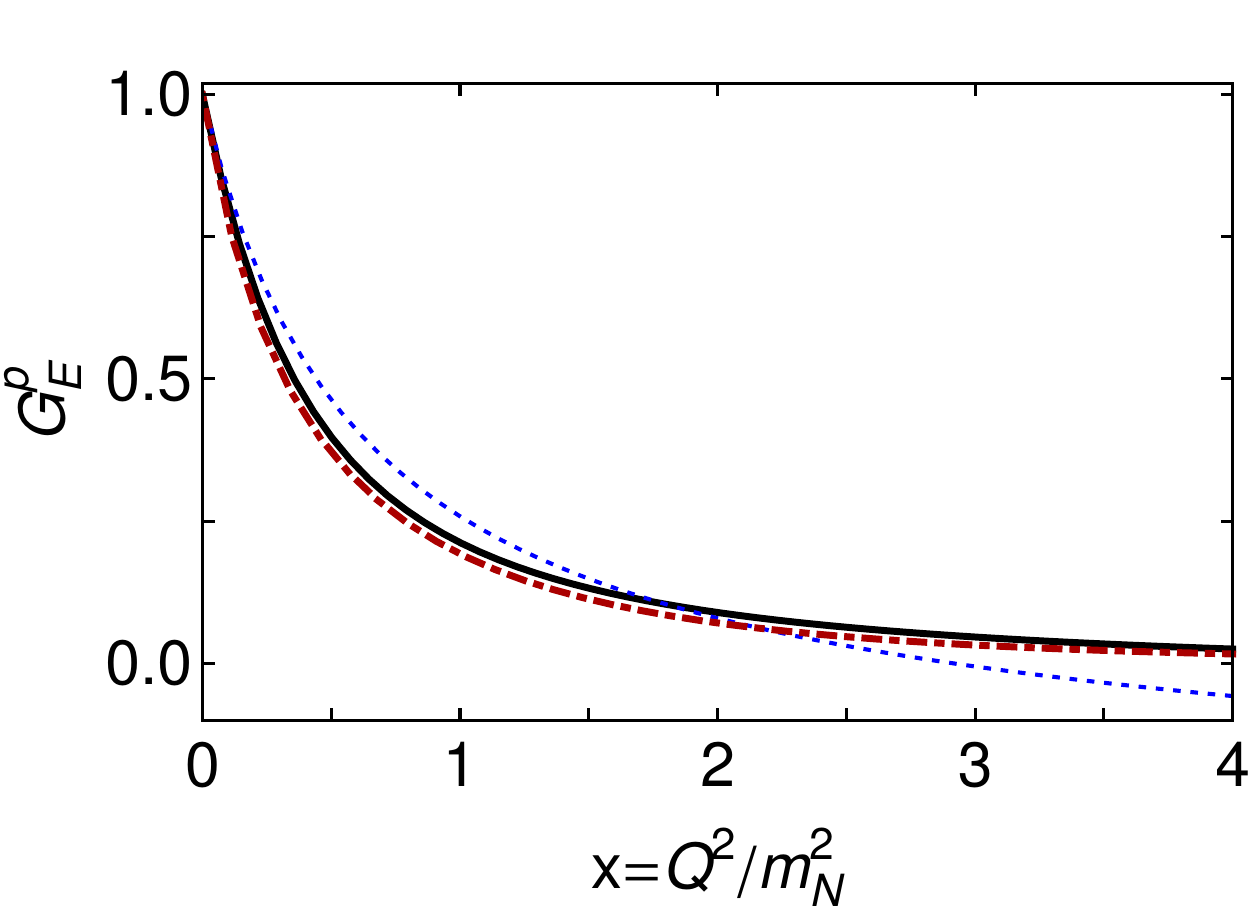} & \hspace*{0.50cm} &
\includegraphics[clip, width=0.45\linewidth]{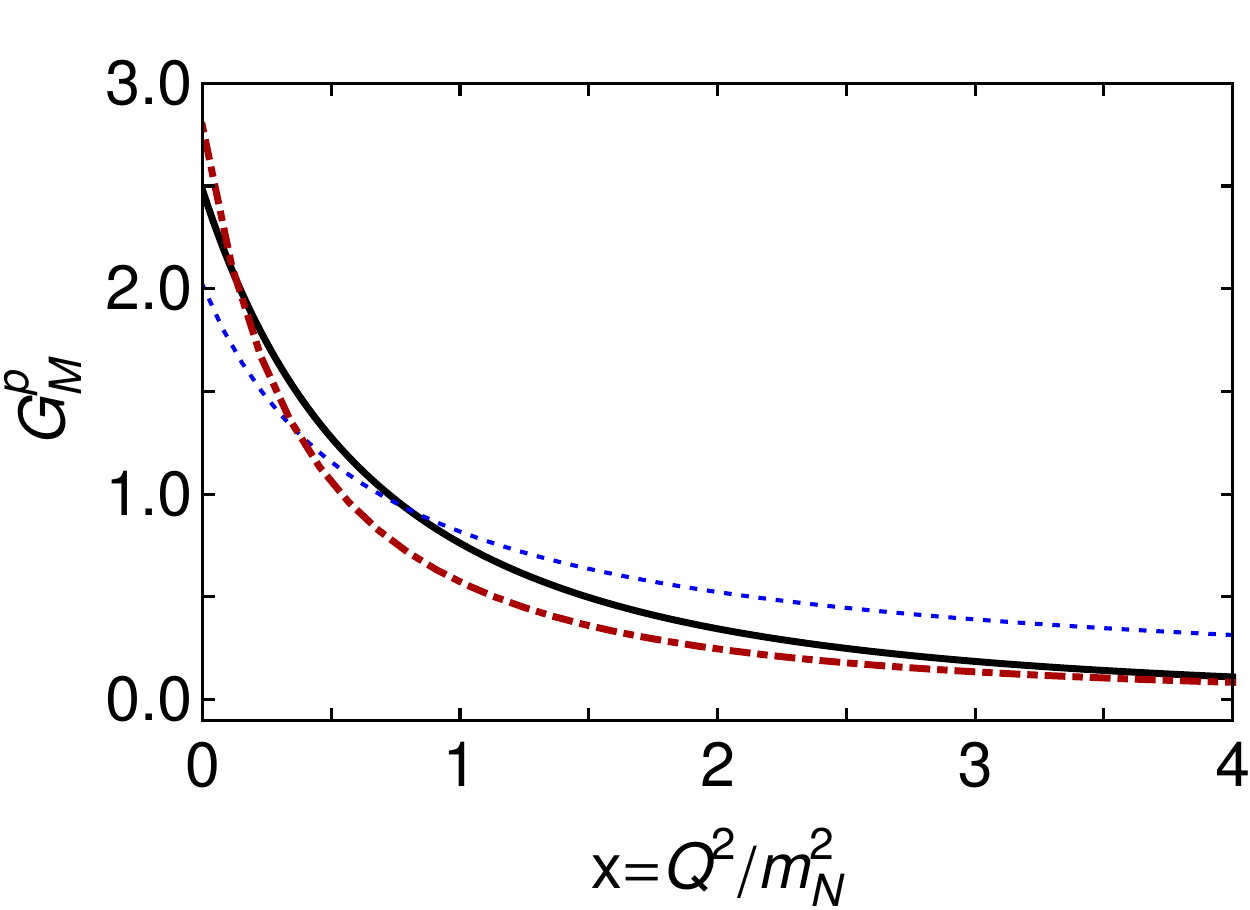} \\[-2ex]
\includegraphics[clip, width=0.45\linewidth]{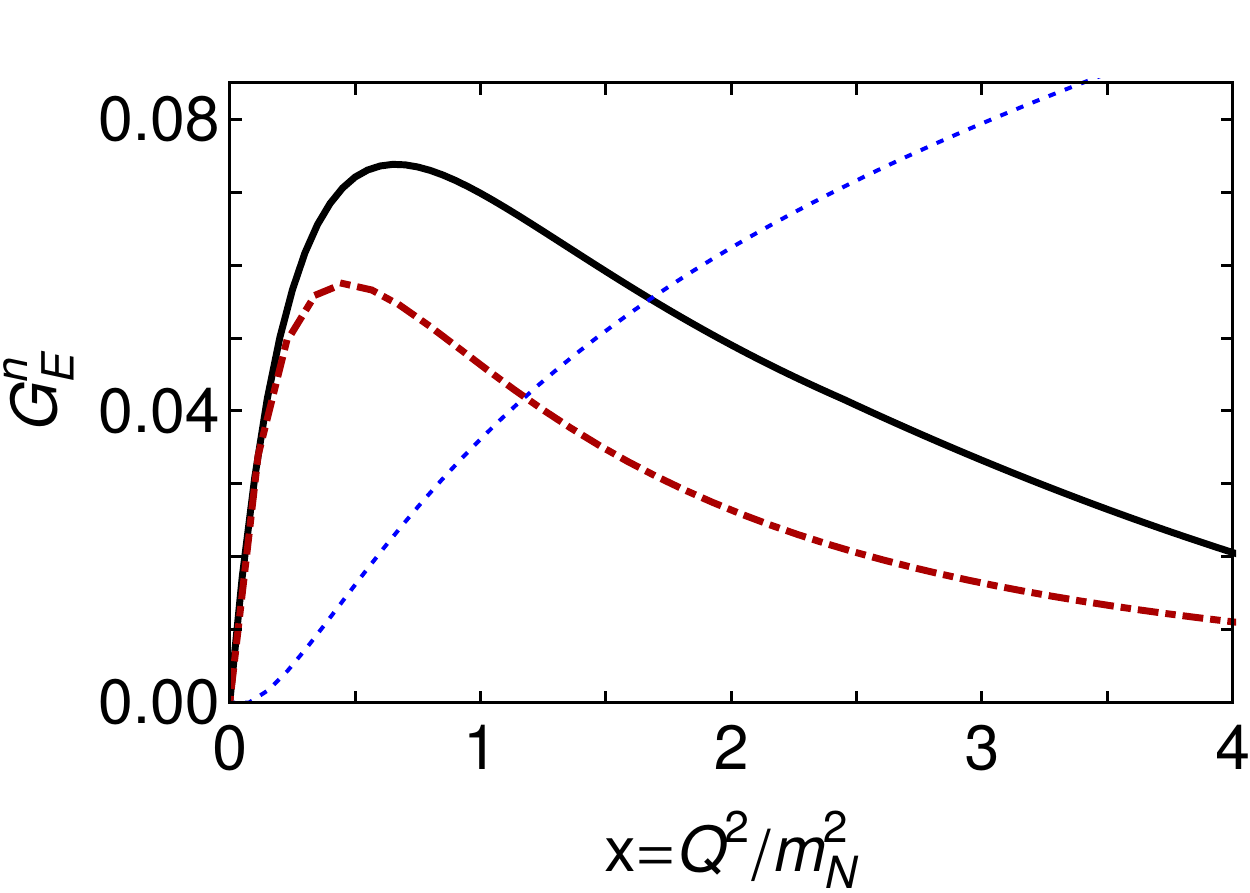} & \hspace*{0.50cm} &
\includegraphics[clip, width=0.45\linewidth]{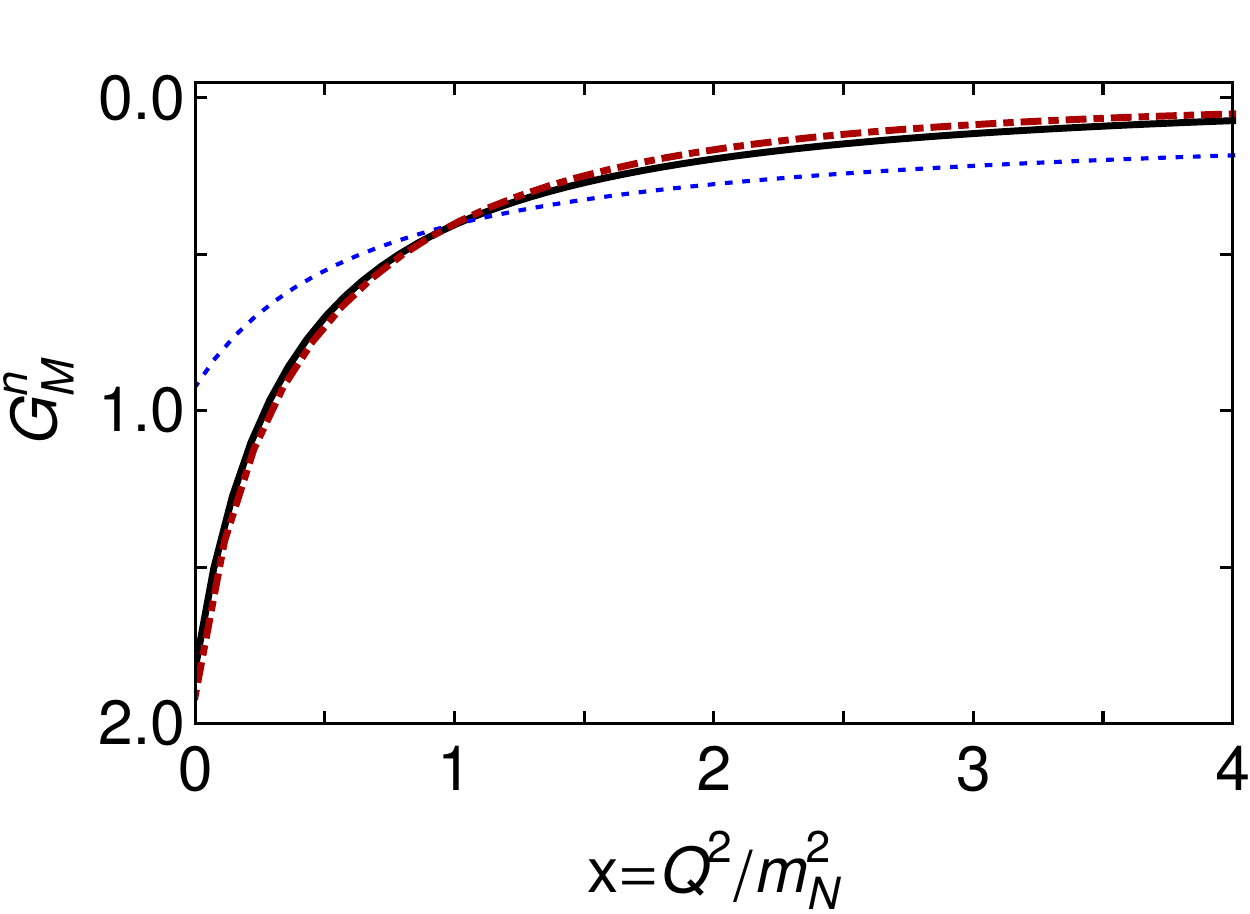}
\put(-65,75){\small QCD-kindred}
\put(-65,60){\small \textcolor{blue}{CI-model}}
\put(-65,45){\small \textcolor{red}{Experiment}}
\end{tabular}
\caption{\label{fig:FFNucleon1}
Proton (top) and neutron (bottom) electromagnetic form factors.
In both rows: {\it left panel} -- Sachs electric; {\it right panel} -- Sachs magnetic.
Curves in all panels: {\it solid, black} -- result obtained using a Faddeev equation kernel and interaction vertices that possess QCD-kindred momentum dependence; viz., a QCD-kindred framework; {\it dotted, blue} -- result obtained with a symmetry preserving treatment of a contact interaction (CI framework)~\cite{Wilson:2011aa}; {\it dot-dashed,red} -- 2004 parametrisation of experimental data~\cite{Kelly:2004hm}.}
\vspace*{-0.40cm}
\end{center}
\end{figure}

Figure~\ref{fig:FFNucleon1} depicts the dressed-quark core Sachs electric and magnetic form factors for the proton and neutron. It is apparent that the QCD-kindred results are in fair agreement with experiment, which is represented by the year-2004 parametrisation in Ref.~\cite{Kelly:2004hm}. Comparisons made with a more recent parametrisation~\cite{Bradford:2006yz} are not materially different. The most notable mismatch appears to be in our description of the neutron electric form factor at low $Q^2$. However, appearances are somewhat deceiving in this case because $G_E^n$ is small on the low-$Q^2$ domain and hence slight differences appear large; moreover, $G_E^n$ is much affected by subdominant effects such as meson-cloud contributions that we have neglected. On the other hand, as was previously observed~\cite{Wilson:2011aa}, form factors obtained via a symmetry-preserving DSE treatment of a contact-interaction are typically too hard. The defects of a contact-interaction are expressed with greatest force in the neutron electric form factor.

\begin{figure}[!t]
\centerline{%
\includegraphics[clip, height=0.25\textheight, width=0.475\textwidth]{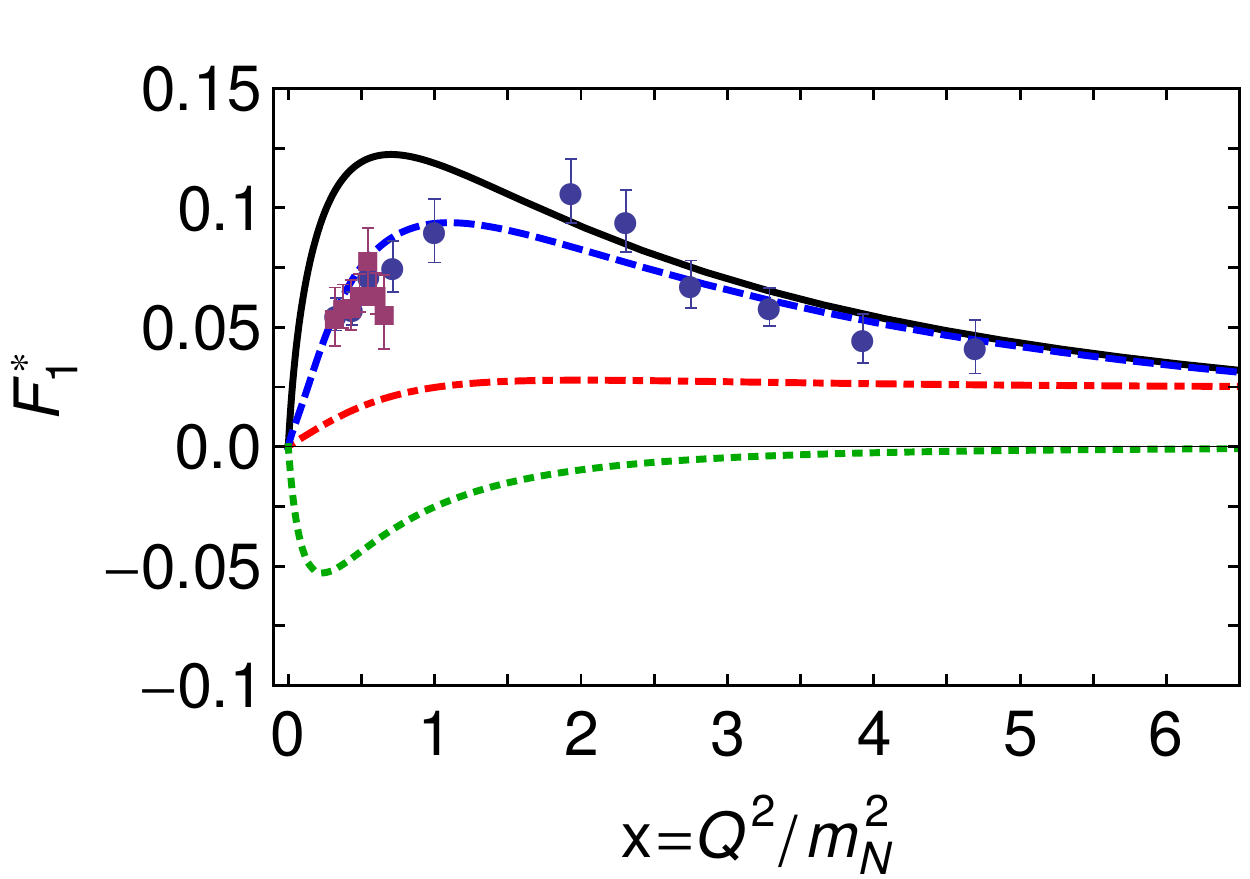}
\hspace*{0.25cm}
\includegraphics[clip, height=0.25\textheight, width=0.475\textwidth]{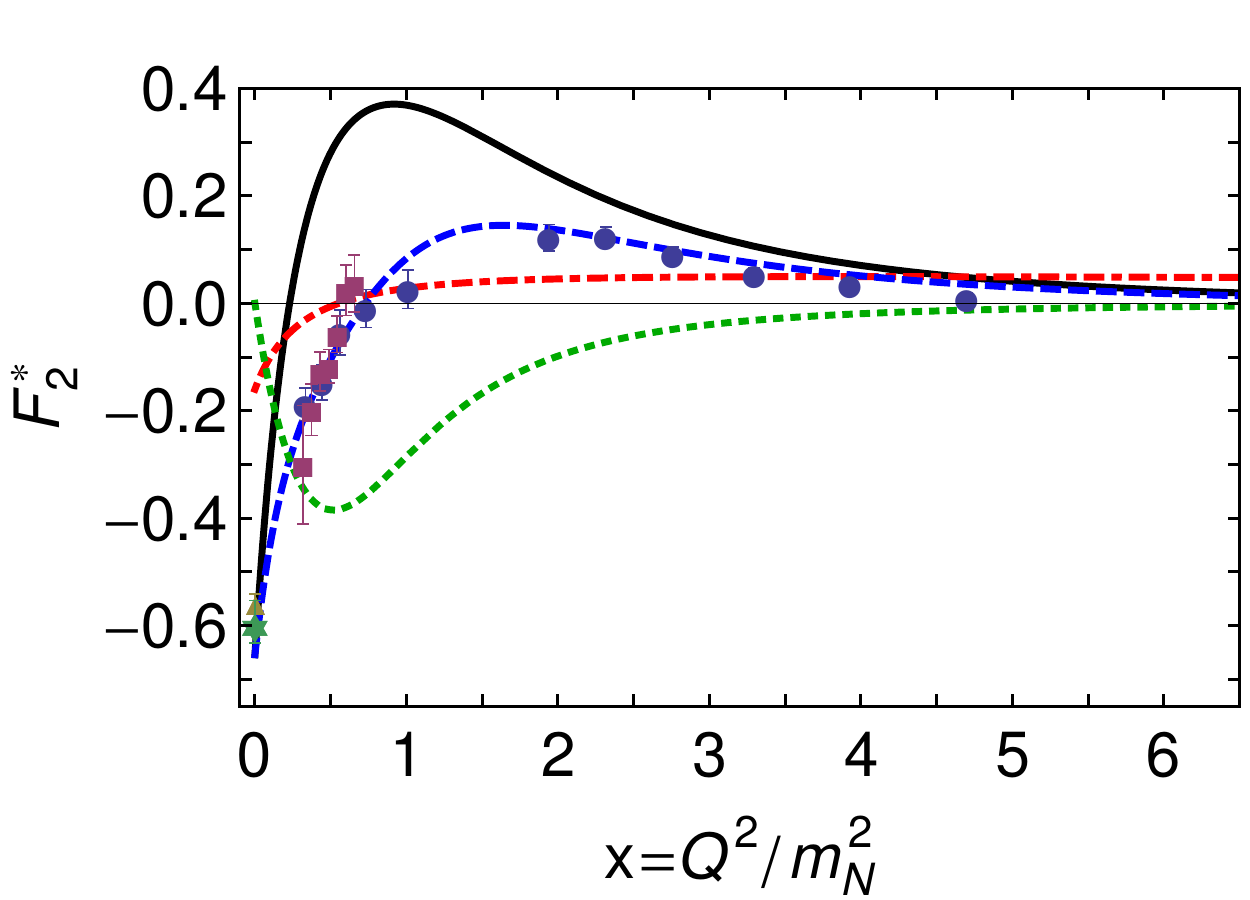}
\put(-60,82){\small QCD-kindred}
\put(-60,70){\small \textcolor{red}{CI-model}}
\put(-60,58){\small \textcolor{blue}{Fit}}
\put(-60,46){\small \textcolor{dgreen}{MB-FSIs}}
}
\caption{\label{fig:NucRop_v2} \emph{Left} -- Dirac transition form factor, $F_{1}^{\ast}(x)$, $x=Q^2/m_N^2$. Solid (black) curve, QCD-kindred prediction; dot-dashed (red) curve, contact-interaction result; dotted (green) curve, inferred meson-cloud contribution; and dashed (blue) curve, anticipated complete result. \emph{Right} -- Pauli transition form factor, $F_{2}^{\ast}(x)$, with same legend. Data in both panels: circles (blue)~\cite{Aznauryan:2009mx}; triangle (gold)~\cite{Dugger:2009pn}; squares (purple)~\cite{Mokeev:2012vsa}; and star (green)~\cite{Tanabashi:2018oca}.
}
\vspace*{-0.40cm}
\end{figure}

The equivalent Dirac and Pauli form factors of the $\gamma^{\ast}p\to R^+$ transition are displayed in Fig.~\ref{fig:NucRop_v2}. The results obtained using QCD-derived propagators and vertices agree with the data on $x\gtrsim 2$. The contact-interaction result simply disagrees both quantitatively and qualitatively with the data. Therefore, experiment is evidently a sensitive tool with which to chart the nature of the quark-quark interaction and hence discriminate between competing theoretical hypotheses. The disagreement between the QCD-kindred result and data on $x\lesssim 2$ is due to meson-cloud contributions that are expected to be important on this domain~\cite{Suzuki:2009nj, Segovia:2015hra, Roberts:2016dnb, Aznauryan:2016wwm, Burkert:2019djo}. An inferred form of that contribution is provided by the dotted (green) curves in Fig.~\ref{fig:NucRop_v2}. They are small already at $x=2$ and vanish rapidly thereafter so that the quark-core prediction remain as the explanation of the data. It is worth to emphasize that the zero crossing in $F_{2}^{\ast}$ is always present but its precise location depends on the meson-cloud estimation.

\begin{figure}[!t]
\includegraphics[clip, height=0.25\textheight, width=0.45\linewidth]{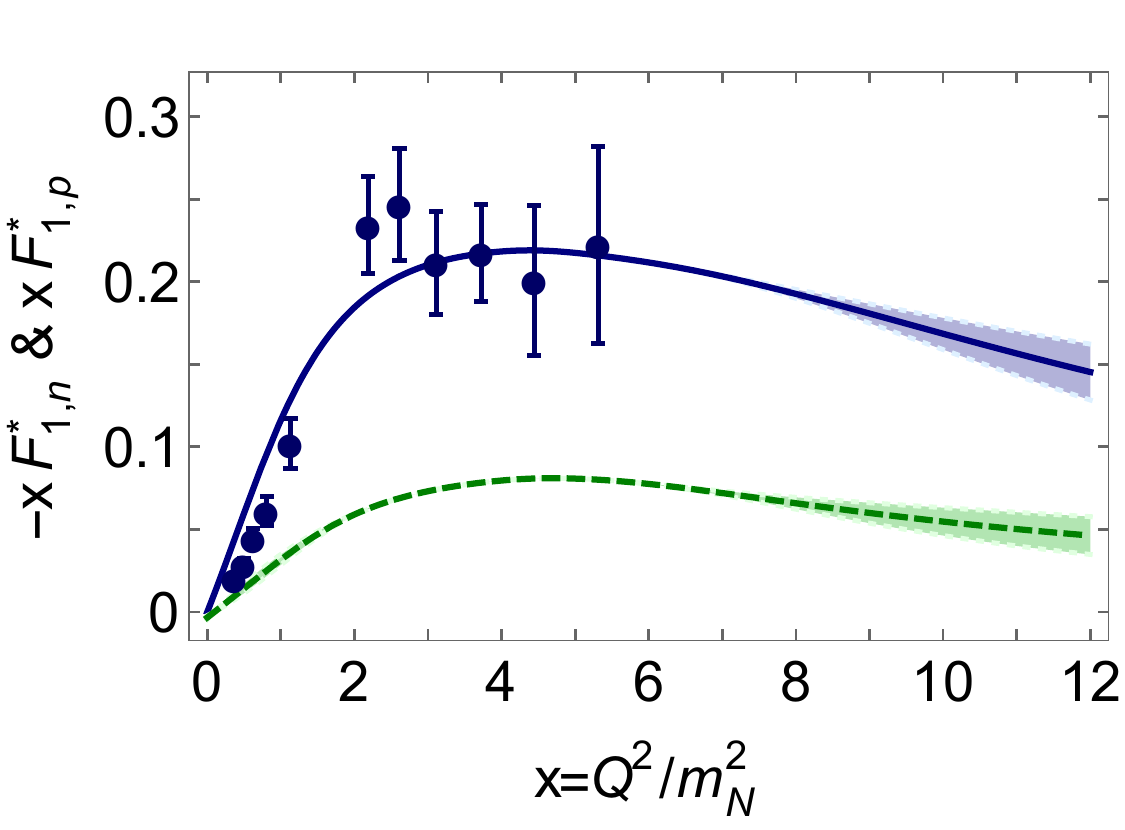}
\hspace*{0.50cm}
\includegraphics[clip, height=0.25\textheight, width=0.45\linewidth]{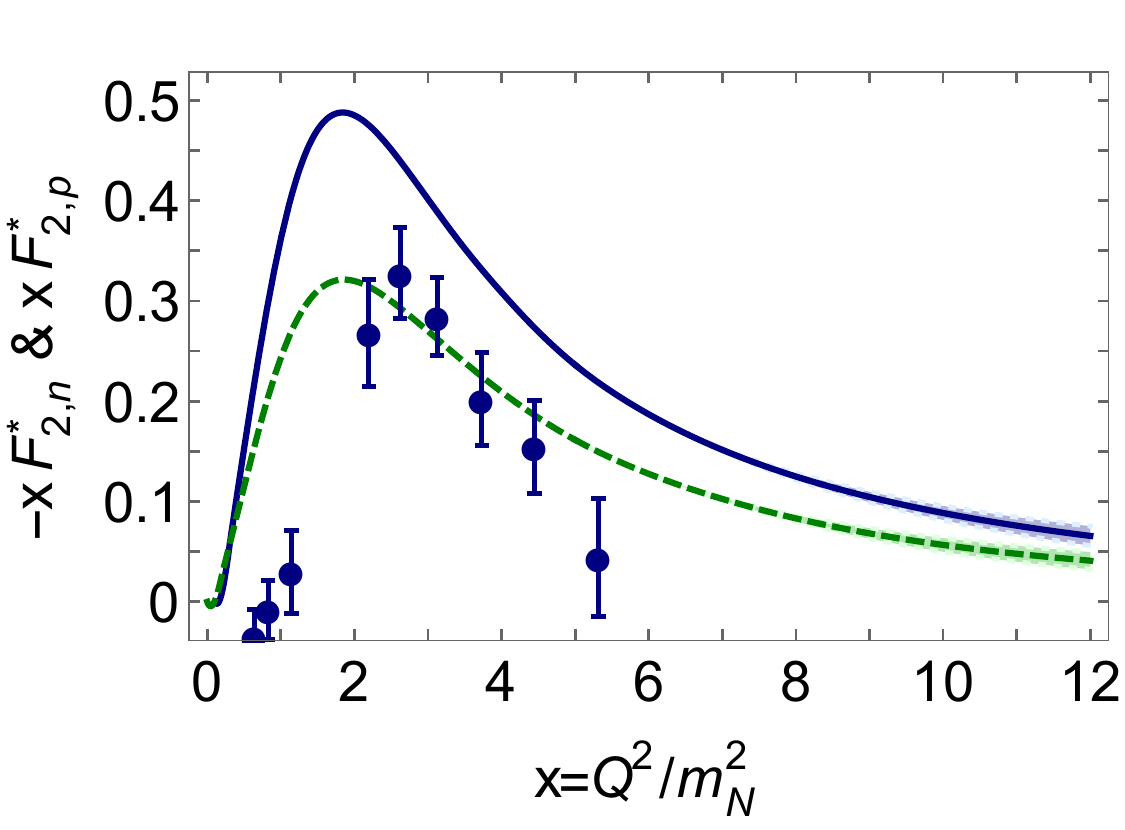}
\caption{\label{figLargeQ2}
Computed $x$-weighted Dirac (left panel) and Pauli (right panel) transition form factors for the reactions $\gamma^\ast\,p\to R^+$ (solid blue curves) and $\gamma^\ast\,n\to R^0$ (dashed green curves). In all cases, the results on $x\in [6,12]$ are projections, obtained via extrapolation of analytic approximations to our results on $x\in [0,6]$ (see Ref.~\cite{Chen:2018nsg} for details). The width of the band associated with a given curve indicates our confidence in the extrapolated value.
Data in both panels are for the charged channel transitions, $F_{1,p}^\ast$ and $F_{2,p}^\ast$: circles (blue)~\cite{Aznauryan:2009mx}. No data currently exist for the neutral channel but they are expected.
}
\vspace*{-0.40cm}
\end{figure}

Finally, since it is anticipated that CLAS~12 detector will deliver data on the Roper-resonance electro-production form factors out to $Q^2 \sim 12 m_N^2$, we depict in Fig.~\ref{figLargeQ2} the $x$-weighted Dirac and Pauli transition form factors for the reactions $\gamma^\ast p \to R^{+}$, $\gamma^\ast n\to R^{0}$ on the domain $0<x<12$.
On the domain depicted, there is no indication of the scaling behaviour expected of the transition form factors: $F^\ast_{1} \sim 1/x^2$, $F^\ast_2 \sim 1/x^3$. Since each dressed-quark in the baryons must roughly share the impulse momentum, $Q$, we expect that such behaviour will only become evident on $x\gtrsim 20$.

%


\section{\label{sec:delta} THE $\gamma^{\ast}p\to \Delta(1232),\,\Delta(1600)$ TRANSITIONS}

The experimental data on $\gamma^\ast p \to \Delta(1232)$ transition are available for $0 \leq Q^2 \lesssim 8\,$GeV$^2$~\cite{Aznauryan:2011qj, Aznauryan:2012ba} and have stimulated much theoretical analysis and speculation about, \emph{inter alia}: the relevance of perturbative QCD (pQCD) to processes involving moderate momentum transfers~\cite{Carlson:1985mm, Pascalutsa:2006up, Aznauryan:2011qj, Aznauryan:2012ba, Eichmann:2011aa}; hadron shape deformation~\cite{Eichmann:2011aa, Alexandrou:2012da, Santopinto:2012nq, Sanchis-Alepuz:2017mir}; and the role that resonance electroproduction experiments can play in exposing nonperturbative aspects of QCD, such as the nature of confinement and dynamical chiral symmetry breaking (DCSB)~\cite{Aznauryan:2012ba}.

Contradicting quark-model predictions~\cite{Capstick:2000qj, Crede:2013sze}, the first positive-parity excitation, $\Delta(1600)\,3/2^+$, lies below the negative parity $\Delta(1700)\,3/2^-$, with the splitting being approximately the same as that in the nucleon sector. This being the case and given the Roper-resonance example above, it is likely that elucidating the nature of the $\Delta(1600)\,3/2^+$-baryon will require both (\emph{i}) data on its electroproduction form factors which extends well beyond the meson-cloud domain and (\emph{ii}) predictions for these form factors to compare with that data. The data exist~\cite{Trivedi:2018rgo, Burkert:2019opk}; and can be analysed with this aim understood.

This section shows our results for the $\gamma^\ast p\to \Delta(1232),\,\Delta(1600)$ transitions, providing comparisons with data and other analyses when available. It is mostly based on Refs.~\cite{Segovia:2013rca, Segovia:2013uga, Segovia:2014aza, Segovia:2016zyc, Chen:2019fzn, Lu:2019bjs} and the interested reader is referred to such references for further details.


\subsection{Transition Current}

Electromagnetic $N\to\Delta$ transitions are described by three form factors~\cite{Jones:1972ky}: magnetic-dipole, $G_M^\ast$; electric quadrupole, $G_E^\ast$; and Coulomb (longitudinal) quadrupole, $G_C^\ast$. They arise through consideration of the transition current:
\begin{equation}
J_{\mu\lambda}(K,Q) =
\Lambda_{+}(P_{f}) R_{\lambda\alpha}(P_{f}) i\gamma_{5} \Gamma_{\alpha\mu}(K,Q) \Lambda_{+}(P_{i}),
\label{eq:JTransition}
\end{equation}
where: $P_{i}$, $P_{f}$ are, respectively, the incoming nucleon and outgoing $\Delta$ momenta, $P_{i}^{2}=-m_{N}^{2}$, $P_{f}^{2}=-m_{\Delta}^{2}$; $Q=P_{f}-P_{i}$ is the incoming photon momentum, $K=(P_{i}+P_{f})/2$; and $\Lambda_{+}(P_{i})$, $\Lambda_{+}(P_{f})$ are, respectively, positive-energy projection operators for the nucleon and $\Delta$, with the Rarita-Schwinger tensor projector $R_{\lambda\alpha}(P_f)$ arising in the latter connection. (See Ref.~\cite{Segovia:2014aza}, Appendix~B.)

In order to succinctly express $\Gamma_{\alpha\mu}(K,Q)$, we define
\begin{equation}
\check K_{\mu}^{\perp} = {\cal T}_{\mu\nu}^{Q} \check{K}_{\nu}
= (\delta_{\mu\nu} - \check{Q}_{\mu} \check{Q}_{\nu}) \check{K}_{\nu},
\end{equation}
with $\check{K}^{2} = 1 = \check{Q}^{2}$, in which case
\begin{equation}
\Gamma_{\alpha\mu}(K,Q) = \mathpzc{k} \left[ \frac{\lambda_m}{2\lambda_{+}}(G_{M}^{\ast} - G_{E}^{\ast})\gamma_{5}
\varepsilon_{\alpha\mu\gamma\delta} \check K_{\gamma}\check{Q}_{\delta} -
G_{E}^{\ast} {\cal T}_{\alpha\gamma}^{Q} {\cal T}_{\gamma\mu}^{K}
- \frac{i\varsigma}{\lambda_m}G_{C}^{\ast}\check{Q}_{\alpha} \check
K^\perp_{\mu}\right],
\label{eq:Gamma2Transition}
\end{equation}
where
$\mathpzc{k} = \sqrt{(3/2)}(1+m_\Delta/m_N)$,
$\varsigma = Q^{2}/[2\Sigma_{\Delta N}]$,
$\lambda_\pm = \varsigma + t_\pm/[2 \Sigma_{\Delta N}]$
with $t_\pm = (m_\Delta \pm m_N)^2$,
$\lambda_m = \sqrt{\lambda_+ \lambda_-}$,
$\Sigma_{\Delta N} = m_\Delta^2 + m_N^2$, $\Delta_{\Delta N} = m_\Delta^2 - m_N^2$.

With a concrete expression for the current in hand, one may obtain the form factors using any three sensibly chosen projection operators, \emph{e.g}.\ with~\cite{Eichmann:2011aa}
\begin{equation}
\label{ProjectionsE}
\mathpzc{t}_{1} = \mathpzc{n}
\frac{\sqrt{\varsigma(1+2\mathpzc{d})}}{\mathpzc{d}-\varsigma}
{\cal T}^{K}_{\mu\nu}\check K^\perp_{\lambda} {\rm tr}
\gamma_{5}J_{\mu\lambda}\gamma_{\nu} \,, \quad
\mathpzc{t}_{2} = \mathpzc{n} \frac{\lambda_{+}}{\lambda_m} {\cal T}^{K}_{\mu\lambda}
{\rm tr} \gamma_{5} J_{\mu \lambda} \,, \quad
\mathpzc{t}_{3} =  3 \mathpzc{n}
\frac{\lambda_+}{\lambda_m}\frac{(1+2\mathpzc{d})}{\mathpzc{d}-\varsigma} \check
K^\perp_{\mu}\check K^\perp_{\lambda} {\rm tr}\gamma_{5}J_{\mu\lambda} \,,
\end{equation}
where $\mathpzc{d}=\Delta_{\Delta N}/[2 \Sigma_{\Delta N}]$,
$\mathpzc{n}= \sqrt{1-4\mathpzc{d}^{2}}/[4i\mathpzc{k}\lambda_m]$), then
\begin{equation}
\label{GMGEGC}
G_{M}^{\ast} = 3 \left[ \mathpzc{t}_{2}+\mathpzc{t}_{1}\right]\,, \quad\quad 
G_{E}^{\ast} = \mathpzc{t}_{2}-\mathpzc{t}_{1}\,, \quad\quad
G_{C}^{\ast} = \mathpzc{t}_{3}.
\end{equation}

The following ratios are often considered in connection with $\gamma^\ast N\to \Delta$ transitions:
\begin{equation}
\label{eqREMSM}
R_{\rm EM} = -\frac{G_E^{\ast}}{G_M^{\ast}}, \quad\quad
R_{\rm SM} = - \frac{|\vec{Q}|}{2 m_\Delta} \frac{G_C^{\ast}}{G_M^{\ast}} = - \frac{\Sigma_{\Delta N}\lambda_m}{2m_\Delta^2} \frac{G_c^\ast}{G_M^\ast} \,.
\end{equation}
Since they are identically zero in $SU(6)$-symmetric constituent-quark models, they can be read as measures of deformation in one or both of the hadrons involved.


\begin{figure}[!t]
\begin{center}
\begin{tabular}{ll}
\includegraphics[clip, width=0.40\textwidth]{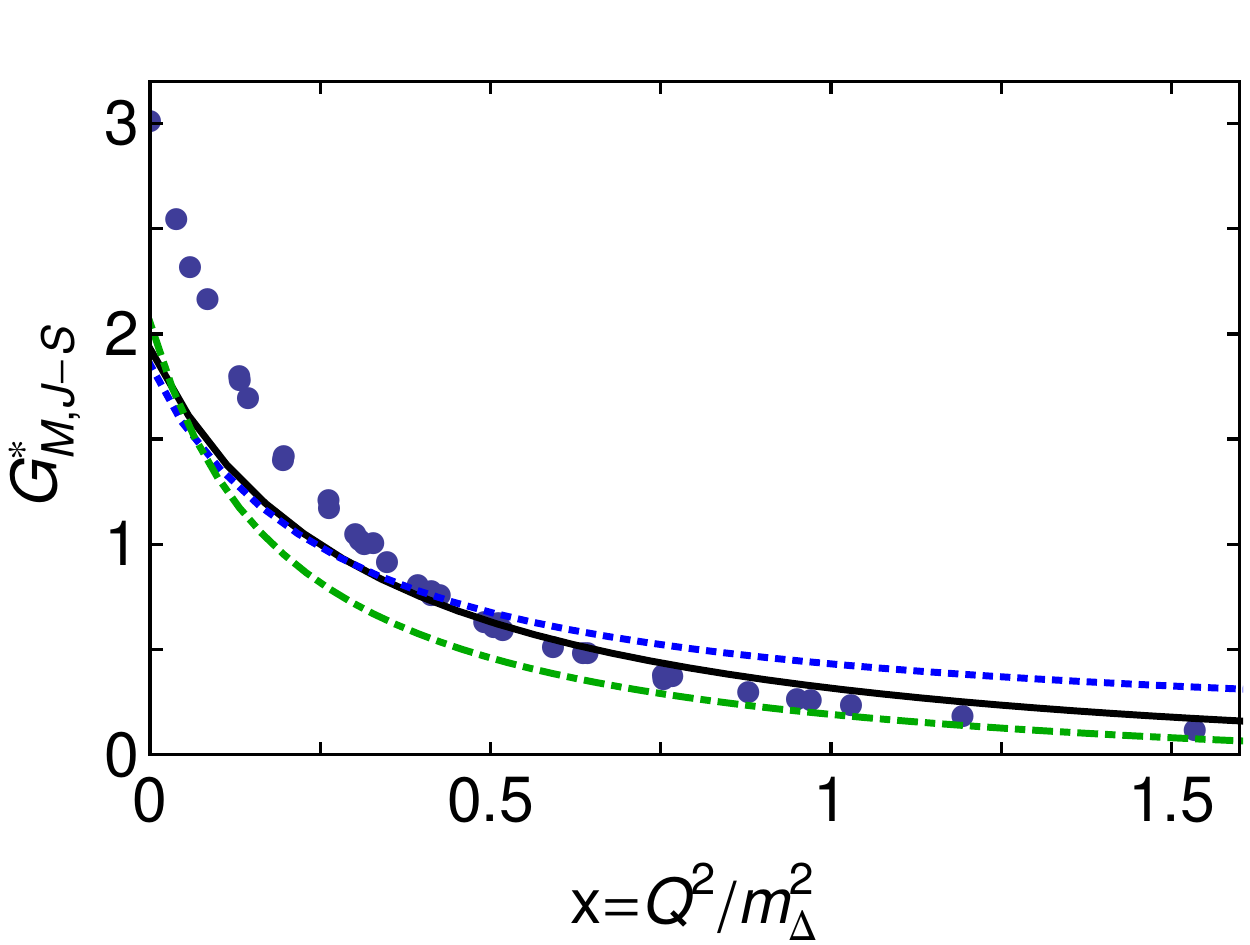}
& 
\hspace*{-0.20cm} 
\includegraphics[clip, height=0.22\textheight, width=0.45\textwidth]{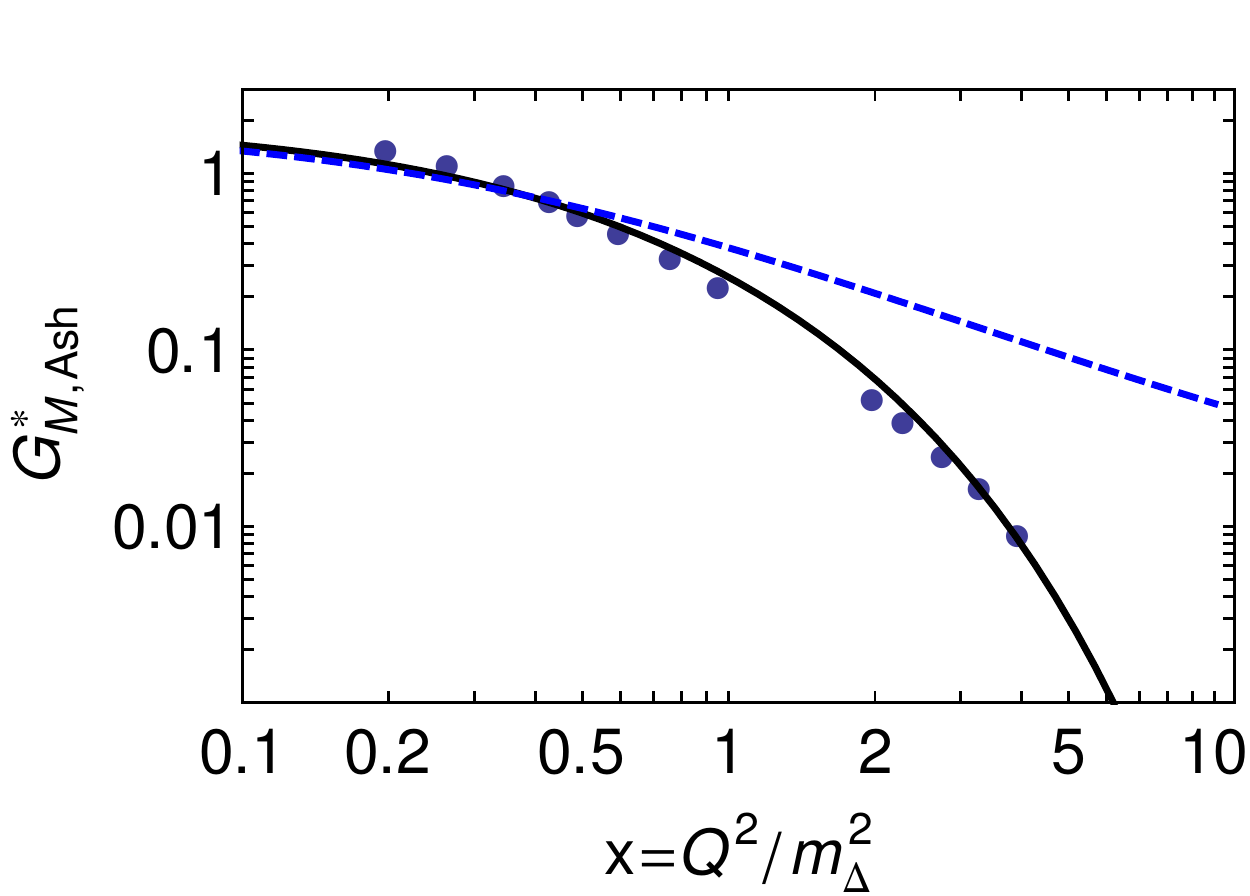} \\[-2ex]
\hspace*{-0.30cm}
\includegraphics[clip, width=0.42\textwidth]{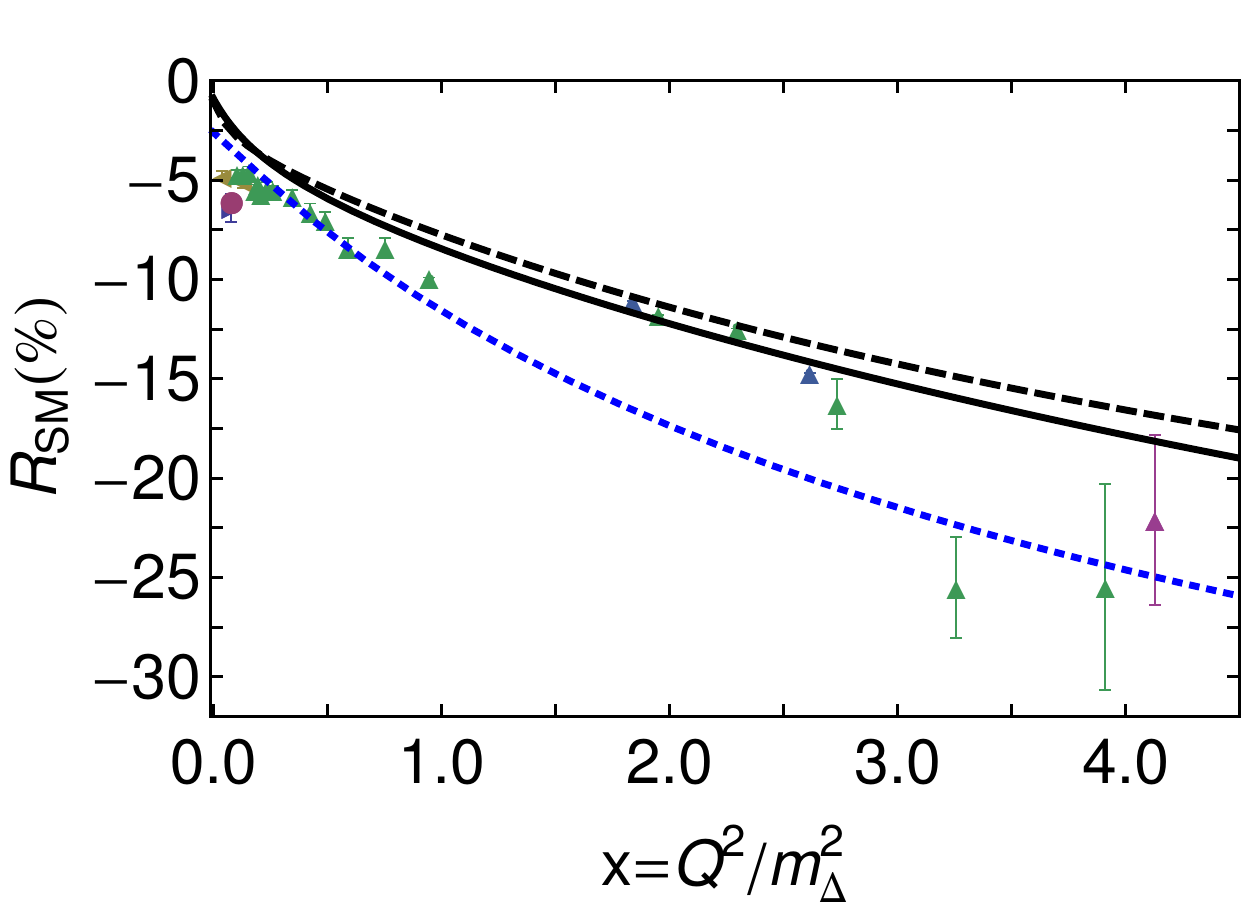}
&
\hspace*{+0.20cm} 
\includegraphics[clip, width=0.42\textwidth]{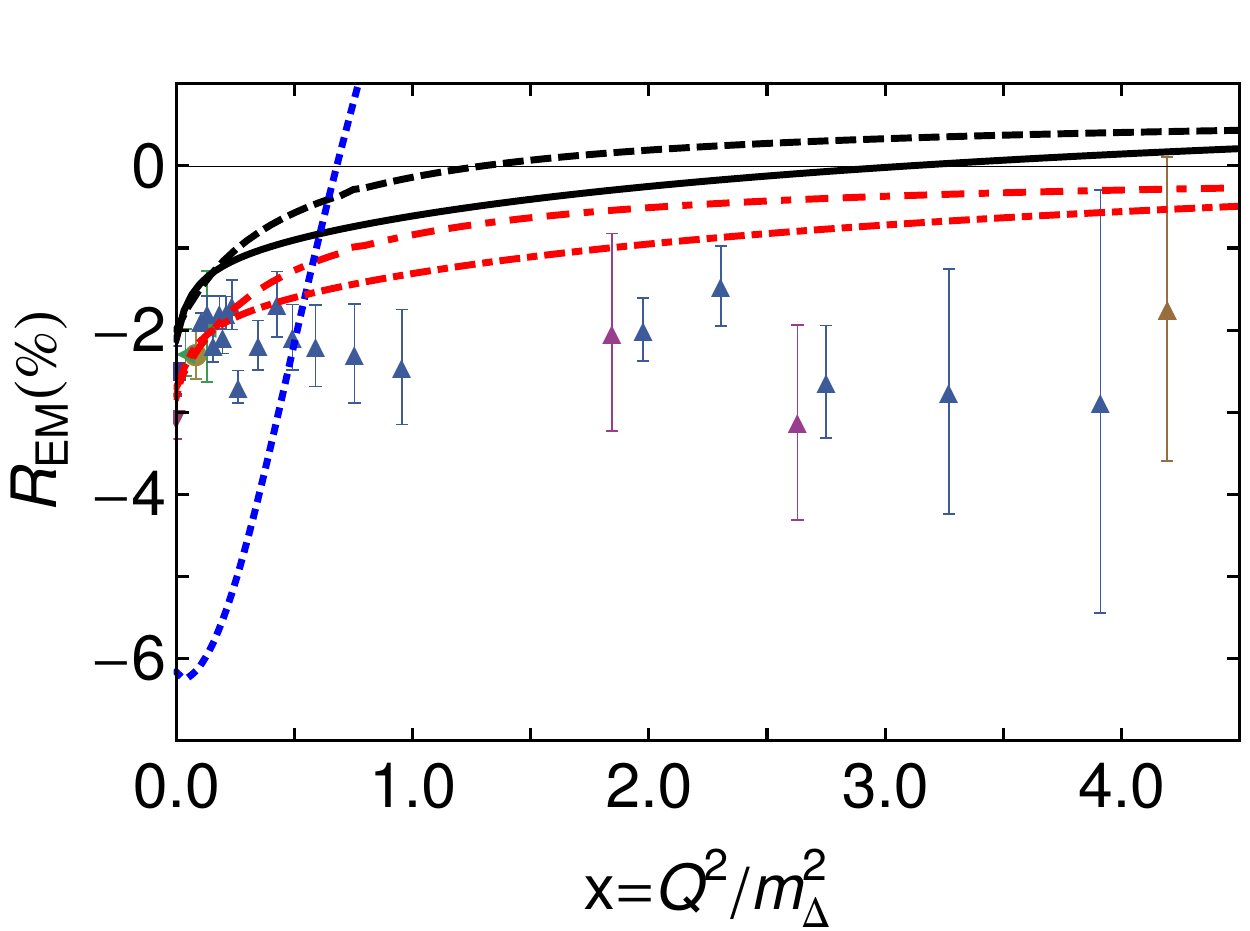}
\end{tabular}
\caption{\label{fig:NucDel} 
\emph{Upper-left panel} -- $G_{M,J-S}^{\ast}$ result obtained with QCD-kindred interaction (solid, black) and with contact-interaction (CI) (dotted, blue); The green dot-dashed curve is the dressed-quark core contribution inferred using Sato-Lee (SL) dynamical meson-exchange model~\protect\cite{JuliaDiaz:2006xt}.
\emph{Upper-right panel} -- $G_{M,Ash}^{\ast}$ result obtained with QCD-kindred interaction (solid, black) and with CI (dotted, blue).
\emph{Lower-left panel} -- $R_{SM}$ prediction of QCD-kindred kernel including dressed-quark anomalous magnetic moment (DqAMM) (black, solid), non-including DqAMM (black, dashed), and CI result (dotted, blue).
\emph{Lower-right panel} -- $R_{EM}$ prediction obtained with QCD-kindred framework (solid, black); same input but without DqAMM (dashed, black); these results renormalized (by a factor of $1.34$) to agree with experiment at $x=0$ (dot-dashed, red - zero at $x\approx 14$; and dot-dash-dashed, red, zero at $x\approx 6$); and CI result (dotted, blue).
The data in the panels are from references that can be found in~\protect\cite{Segovia:2014aza}.
}
\vspace*{-0.40cm}
\end{center}
\end{figure}

\subsection{Form Factors}

The upper-left panel of Fig.~\ref{fig:NucDel} displays the magnetic transition form factor in the Jones-Scadron convention. Both QCD-kindred and CI results agree with the data on $x\gtrsim 0.4$. On the other hand, both curves disagree markedly with the data at infrared momenta. This is related with the fact that Dyson-Schwinger Equations computations ignore meson-cloud effects. The similarity between our curves and the bare result determined using the Sato-Lee (SL) dynamical meson-exchange model~\cite{JuliaDiaz:2006xt} confirms the former assertion.

The upper-right panel of Fig.~\ref{fig:NucDel} shows the magnetic transition form factor in the Ash convention. This is because experimental results on this form factor have been traditionally presented using such convention. One can see that the normalized QCD-kindred curve is in fair agreement with the data, indicating that the Ash form factor falls unexpectedly faster than a dipole form mainly for two reasons: (i) meson-cloud effects provide up-to $35\%$ of the form factor for $x \lesssim 2$; (ii) the additional kinematic factor $\sim 1/\sqrt{Q^2}$ that appears between Ash and Jones-Scadron conventions provides material damping for $x\gtrsim 2$ (see Ref.~\cite{Segovia:2014aza} for details on this aspect). 

The lower-left panel of Fig.~\ref{fig:NucDel} displays the Coulomb quadrupole ratio. Our results computed using either the QCD-kindred or the CI formalism are broadly consistent with available data. This shows that even a contact-interaction can produce correlations between dressed-quarks within Faddeev wave-functions and related features in the current that are comparable in size with those observed empirically. Moreover, suppressing the dressed-quark anomalous magnetic moment (DqAMM) in the transition current has little impact. These remarks highlight that $R_{SM}$ is not particularly sensitive to details of the Faddeev kernel and transition current.

The lower-right panel in Fig.~\ref{fig:NucDel} shows that $R_{\rm EM}$ is a particularly sensitive measure of diquark and orbital angular momentum correlations. The contact-interaction result is negative at low photon virtualities, it crosses zero at an experimentally accessible transfer momentum and then increases with x in order to reach the pQCD limit. On the other hand, we have presented four variants of the QCD-kindred result, which differ primarily in the location of the zero that is a feature of this ratio in all cases we have considered. The inclusion of a DqAMM shifts the zero to a larger value of $x$. Given the uniformly small value of this ratio and its sensitivity to the DqAMM, we judge that meson-cloud effects must play a large role on the entire domain that is currently accessible to experiment.

\begin{figure}[t]
\includegraphics[clip, height=0.20\textheight, width=0.33\textwidth]{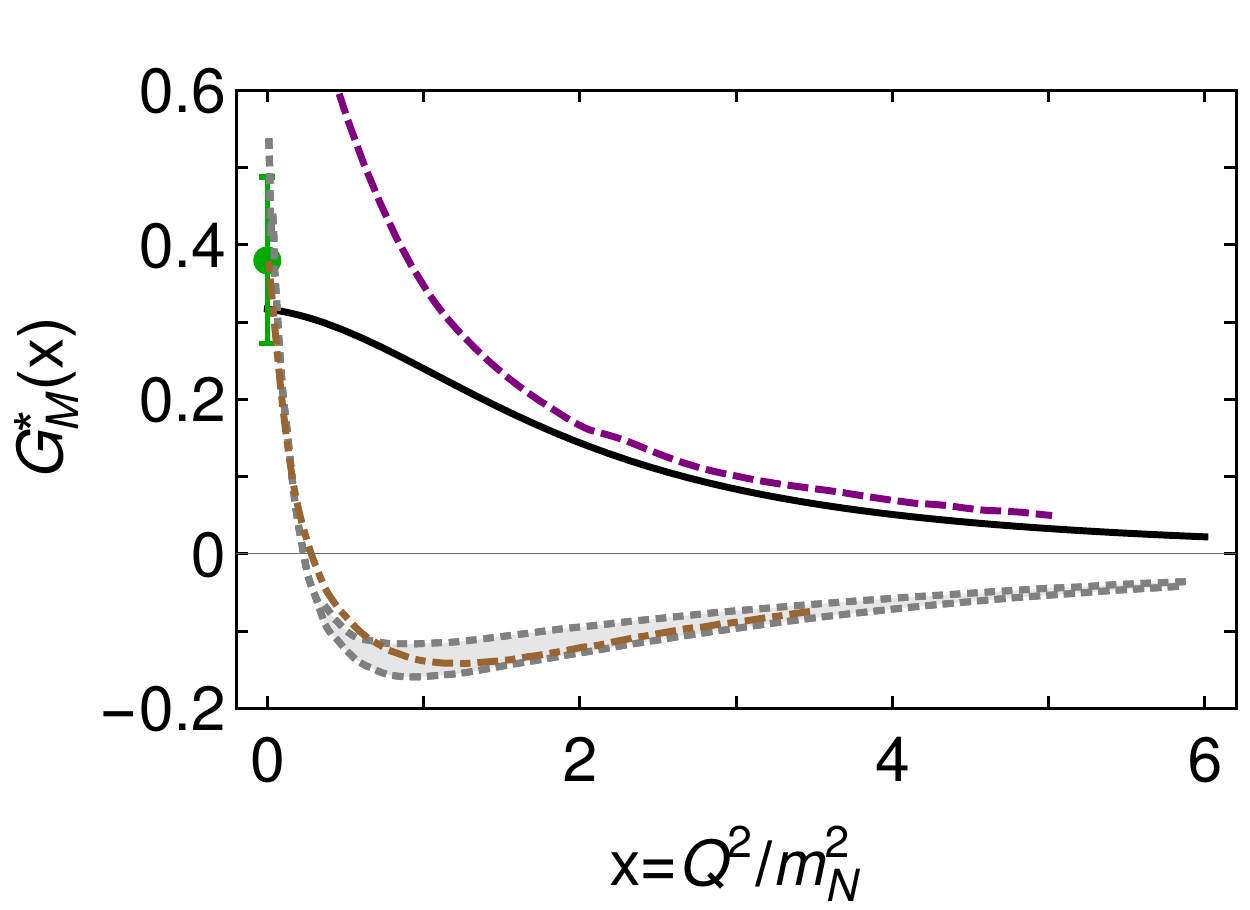}
\includegraphics[clip, height=0.20\textheight, width=0.33\textwidth]{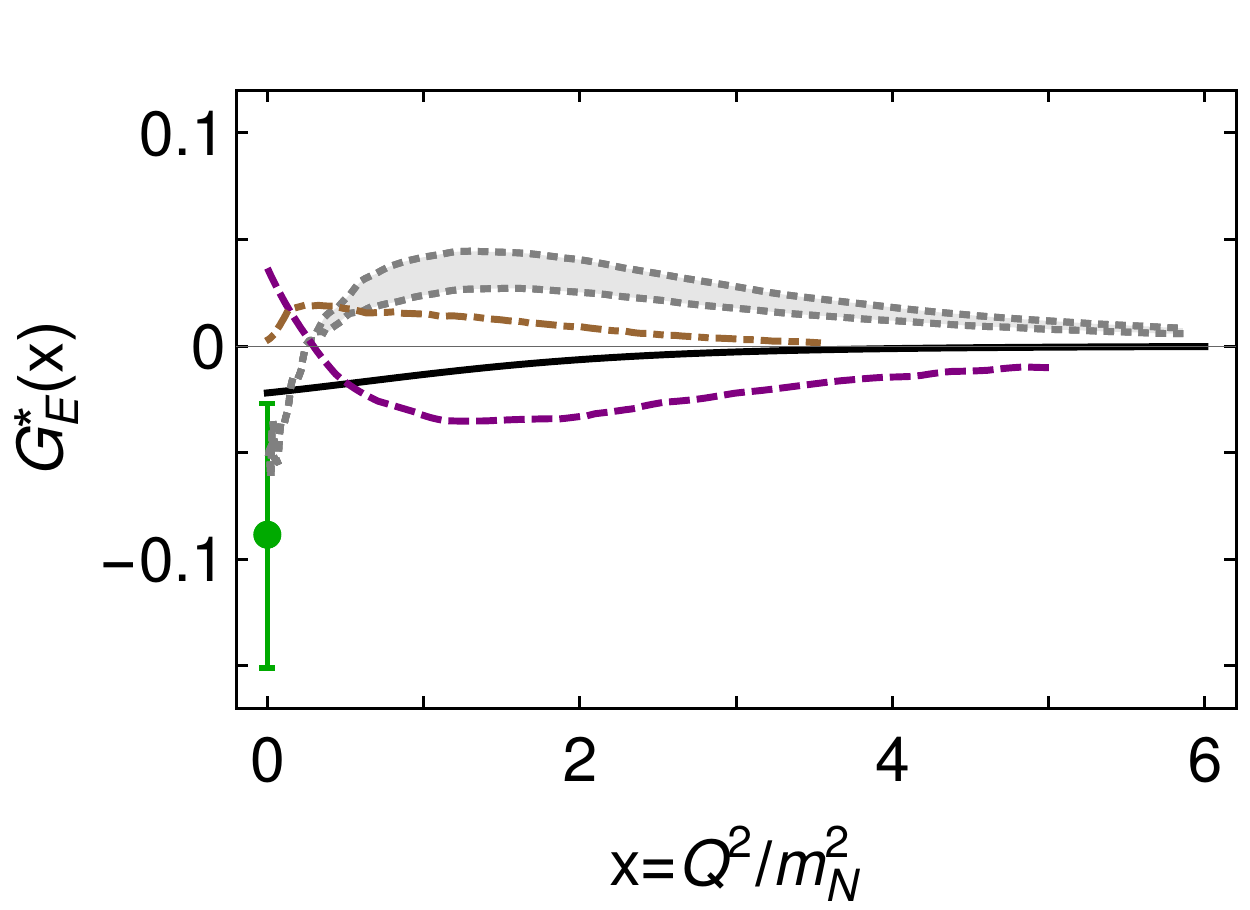}
\includegraphics[clip, height=0.20\textheight, width=0.33\textwidth]{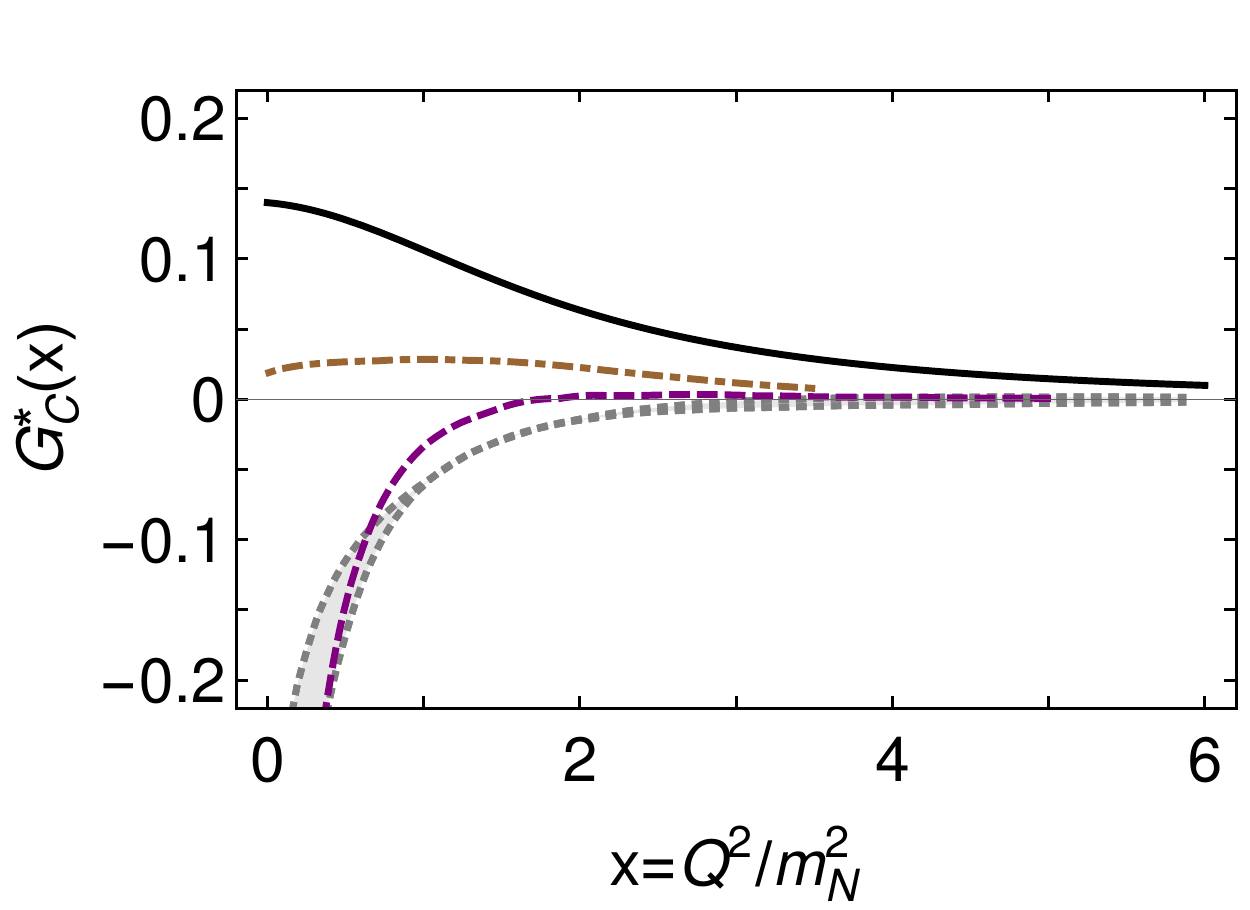}
\\[-2ex]
\includegraphics[clip, height=0.20\textheight, width=0.33\textwidth]{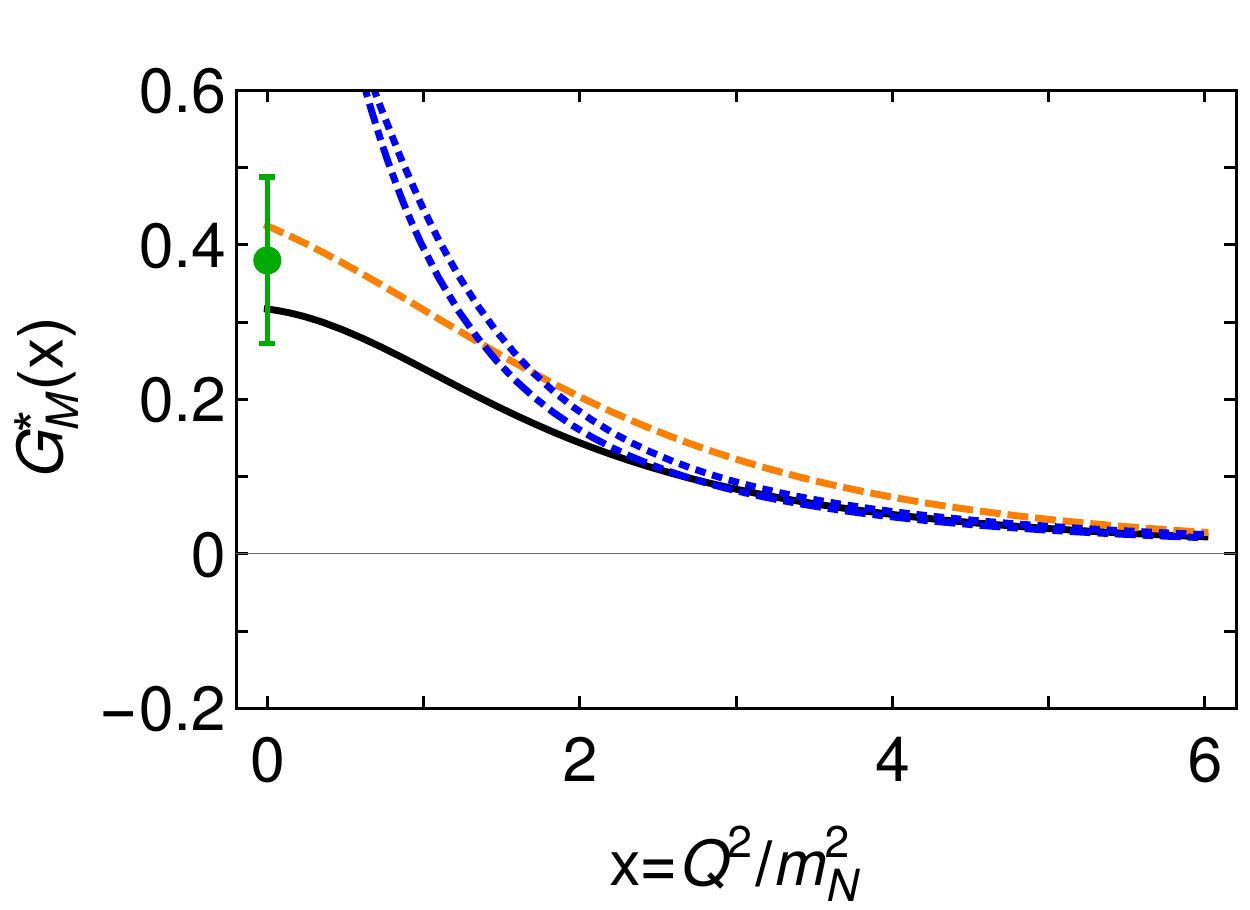}
\includegraphics[clip, height=0.20\textheight, width=0.33\textwidth]{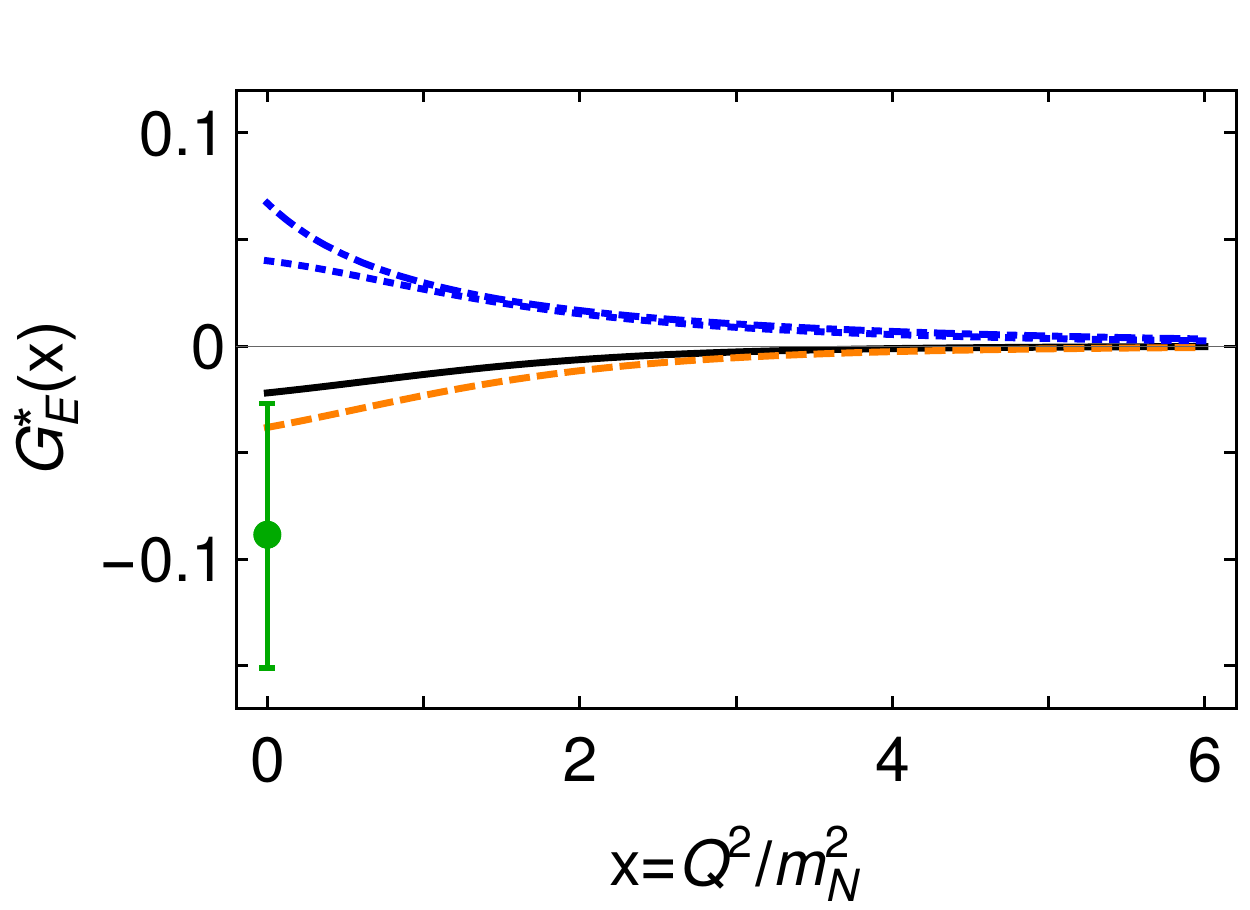}
\includegraphics[clip, height=0.20\textheight, width=0.33\textwidth]{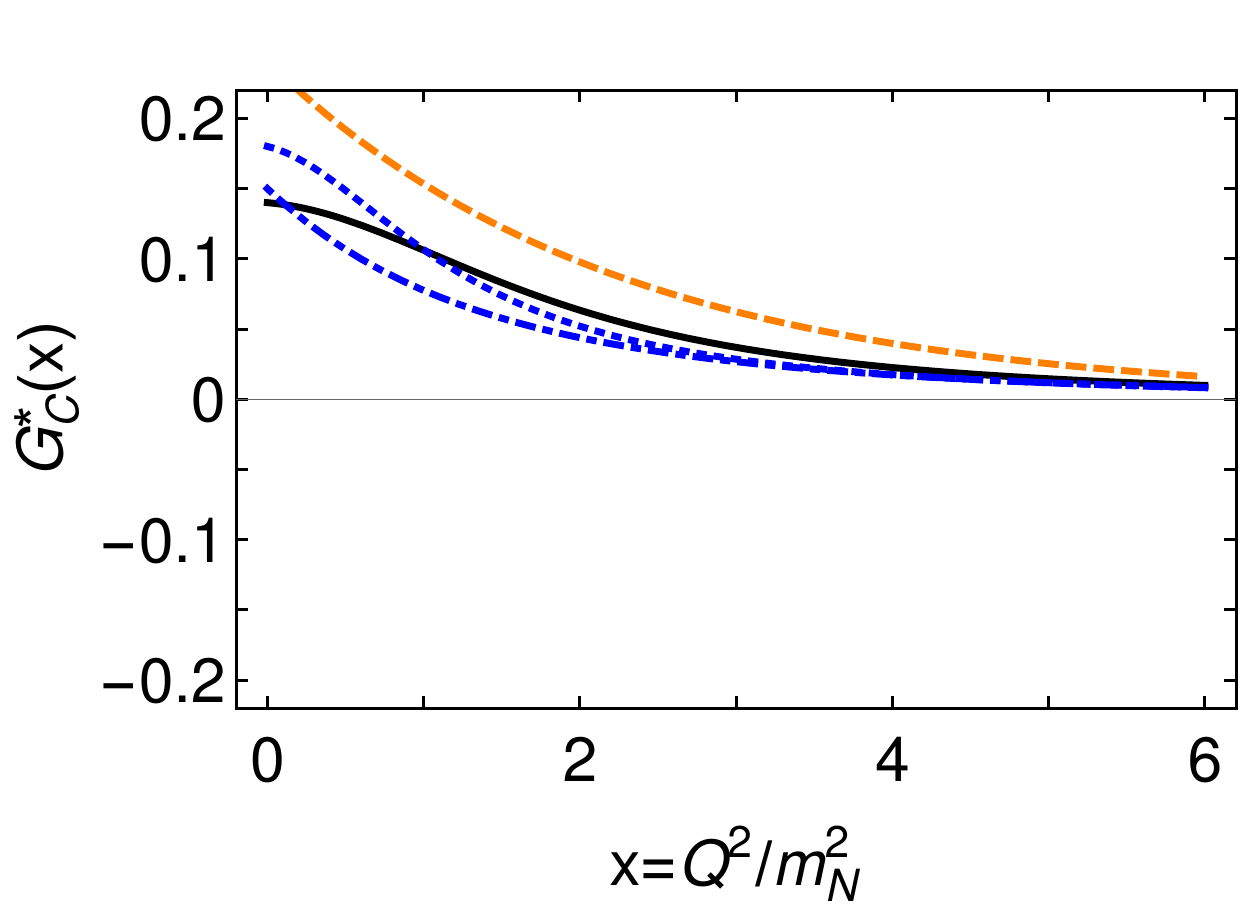}
\caption{\label{D1600TFFs} 
\emph{Left panels} -- Magnetic dipole $\gamma^\ast p\to \Delta^+(1600)$ transition form factor; \emph{middle} -- electric quadrupole; and \emph{right}:  Coulomb quadrupole.
Data from Ref.~\cite{Tanabashi:2018oca}; and the conventions of Ref.~\cite{Jones:1972ky} are employed.
Panels on the top: solid (black) curve, complete result; shaded (grey) band, light-front relativistic Hamiltonian dynamics (LFRHD)~\cite{Capstick:1994ne};
dot-dashed (brown) curve, light-front relativistic quark model (LFRQM) with unmixed wave functions~\cite{Aznauryan:2015zta}; and dashed (purple) curve, LFRQM with configuration mixing~\cite{Aznauryan:2016wwm}.
Panels on the bottom: solid (black) curve, complete result; dotted (blue) curve, both the proton and $\Delta(1600)$ are reduced to $S$-wave states; Dot-dashed (blue) curve, result obtained when $\Delta(1600)$ is reduced to $S$-wave state; dashed (orange) curve, obtained by enhancing proton's axial-vector diquark content.
}
\vspace*{-0.40cm}
\end{figure}

Predictions for the $\gamma^\ast p\to \Delta^+(1600)$ transition form factors are displayed in Fig.~\ref{D1600TFFs}. Empirical results are only available at the real-photon point for two of the three form factors: $G_M^\ast(Q^2=0)$, $G_E^\ast(Q^2=0)$. Evidently, the quark model results -- (shaded grey band)~\cite{Capstick:1994ne}, dot-dashed (brown) curve~\cite{Aznauryan:2015zta} and dashed (purple) curve~\cite{Aznauryan:2016wwm}) -- are very sensitive to the wave functions employed for the initial and final states.  Furthermore, inclusion of relativistic effects has a sizeable impact on transitions to positive-parity excited states~\cite{Capstick:1994ne}.

Our prediction is the solid (black) curve in each panel of Fig.~\ref{D1600TFFs}. In this instance, every transition form factor is of unique sign on the domain displayed. Notably, the mismatches with the empirical results for $G_M^\ast(Q^2=0)$, $G_E^\ast(Q^2=0)$ are commensurate in relative sizes with those in the $\Delta(1232)$ case, suggesting that MB\,FSIs are of similar importance in both channels.

One can mimic some effects of a meson cloud by modifying the axial-vector diquark content of the participating hadrons.  Accordingly, to illustrate the potential impact of MB\,FSIs, we computed the transition form factors using an enhanced axial-vector diquark content in the proton.  This was achieved by setting $m_{1^+} = m_{0^+} = 0.85\,$GeV, values with which the proton's mass is practically unchanged. The procedure produced the dashed (orange) curves in the bottom panels of Fig.~\ref{D1600TFFs}; better aligning the $x\simeq 0$ results with experiment and suggesting thereby that MB\,FSIs will improve our predictions.

The dotted (blue) curve in the bottom panels of Fig.~\ref{D1600TFFs} is the result obtained when only rest-frame $S$-wave components are retained in the wave functions of the proton and $\Delta(1600)$-baryon; and the dot-dashed (blue) curve is that computed with a complete proton wave function and a $S$-wave-projected $\Delta(1600)$. Once again, the higher partial-waves have a visible impact on all form factors, with $G_E^\ast$ being most affected.

In the near term future, the electro-excitation $N\to \Delta(1600)\frac{3}{2}^+$ amplitudes will become available at photon virtualities $2.0\,\text{GeV}^2 < Q^2 < 5.0\,\text{GeV}^2$ from analysis of recent CLAS results on $\pi^+ \pi^- p$ electro-production off proton~\cite{Isupov:2017lnd, Trivedi:2018rgo}. They will allow us to test the continuum QCD expectations on the transition $N\to \Delta(1600)\frac{3}{2}^+$ form factors. The expected experimental results will be of particular importance in order to check universality or environmental sensitivity of the dressed quark mass function for the nucleon and Delta radial excitations.


\section{\label{sec:conclusions} CONCLUSIONS}

We have shown recent calculations of $\gamma^\ast p \to N(940),\,N(1440)$ and $\gamma^\ast p \to \Delta(1232),\,\Delta(1600)$ transition form factors consistent with a quark-diquark approximation to the Poincar\'e-covariant three-body bound-state problem in relativistic quantum field theory. Crucially, the diquark correlations are non-pointlike and fully-dynamical, and the Faddeev kernel ensures that every valence-quark participates actively in all diquark correlations to the fullest extent allowed by kinematics and symmetries. Moreover, each dressed-quark is characterised by a non-perturbatively generated running mass function, expressing a signature consequence of dynamical chiral symmetry breaking in the Standard Model.

Amongst our results, the following are of particular interest: assuming that the first excited state of the nucleon is the so-called Roper resonance, we compare with experiment our computation of the equivalent Dirac and Pauli form factors of the $\gamma^\ast p \to R^+$ reaction and observe that, while the mismatch in the domain of $Q^2\lesssim 2 m_N^2$ may plausibly be attributed to meson-cloud effects, the agreement on $Q^2 \gtrsim 2 m_N^2$ owes fundamentally to the QCD-derived momentum-dependence of the propagators and vertices employed in solving the bound-state and scattering problems.

In connection with the $\gamma^{\ast}p\to \Delta(1232)^+$ transition, the momentum-dependence of the magnetic transition form factor in the Jones-Scadron convention matches that of the nucleon once the momentum transfer is high enough to pierce the meson-cloud. And the electric quadrupole ratio is a keen measure of diquark and orbital angular momentum correlations; its zero crossing, obscured by meson-cloud effects, could be on the domain currently accessible by experiment.

Our predictions for the $\gamma^\ast p \to \Delta^+(1600)$ magnetic dipole and electric quadrupole transition form factors are consistent with the empirical values at the real photon point. In view of the actual discrepancy between theoretical predictions, which basically differ on the assumed $\Delta(1600)$'s wave function, one can conclude that data on the $\gamma^\ast p\to \Delta^+(1600)$ transition form factors will be sensitive to the structure and deformation of the $\Delta^+(1600)$.

\vspace*{-0.10cm}

\begin{acknowledgments}
We are grateful for constructive comments from A. Bashir, D. Binosi, L. Chang, G. Eichmann, C. Fischer, R. W. Gothe, V. Mokeev, S.-X. Qin, J. Rodr\'iguez-Quintero, S. M. Schmidt, F. de Soto, F. Wang and H.-S. Zong.  
Work supported by:
Ministerio de Economia Industria y Competitividad (MINECO), under grant no.\,FPA2017-86380-P; 
Helmholtz International Center for FAIR within the LOEWE program of the State of Hesse and by the DFG grant FI 970/11-1;
National Natural Science Foundation of China, grant nos.\ 11535005, 11805097;
Jiangsu Province Natural Science Foundation grant no.\ BK20180323;
Jiangsu Province \emph{Hundred Talents Plan for Professionals};
and U.S.\ Department of Energy, Office of Science, Office of Nuclear Physics, under contract no.\,DE-AC02-06CH11357.
\end{acknowledgments}

\vspace*{-0.60cm}

\bibliography{Segovia-Jorge-Proceeding}

\end{document}